\documentclass[10pt,conference]{ieeeconf} 

\usepackage{epsfig} 
\usepackage{graphicx}
\usepackage{amsmath, amsfonts}
\usepackage[in]{fullpage}
\usepackage[small, bf]{caption}
\usepackage[normalem]{ulem}
\usepackage[margin=1in]{geometry}
\usepackage{floatrow}
\newfloatcommand{capbtabbox}{table}[][\FBwidth]
\usepackage[sorting=none]{biblatex}
\usepackage{booktabs}
\usepackage[utf8]{inputenc}
\usepackage{multirow} 
\usepackage{array}
\newcolumntype{C}[1]{>{\centering\let\newline\\\arraybackslash\hspace{0pt}}m{#1}}

\newcommand{\fig}[1]{Fig.~\ref{figure:#1}}
\newcommand{\figs}[2]{Figs.~\ref{figure:#1}~and~\ref{figure:#2}}
\newcommand{\figst}[3]{Figs.~\ref{figure:#1},~\ref{figure:#2}~and~\ref{figure:#3}}
\newcommand{\figur}[1]{Figure~\ref{figure:#1}}

\newcommand{\sect}[1]{Section~\ref{section:#1}}

\newcommand{\be}{\begin{equation}}
\newcommand{\ee}{\end{equation}}
\newcommand{\bea}{\begin{eqnarray}}
\newcommand{\eea}{\end{eqnarray}}
\newcommand{\beau}{\begin{eqnarray*}}
\newcommand{\eeau}{\end{eqnarray*}}
\newcommand{\bed}{\begin{displaymath}}
\newcommand{\eed}{\end{displaymath}}

\newcommand{\fev}{\textsc{fev{\ensuremath{_1}}}}
\newcommand{\fvc}{\textsc{fvc}}
\newcommand{\fevf}{\textsc{fev{\ensuremath{_1}}/fvc}}
\newcommand{\fit}{\textsc{fit}}
\newcommand{\tv}{\textsc{tv}}
\newcommand{\rr}{\textsc{rr}}

\begin{document}

\title{Inferring COPD Severity from Tidal Breathing}



\author{Kofi Odame, Graham Atkins, Maria Nyamukuru, Katherine Fearon}

\maketitle

\begin{abstract}
Objective: To develop an algorithm that can infer the \emph{severity level} of a COPD patient's airflow limitation from tidal breathing data that is collected by a wearable device. 

Methods: Data was collected from 25 single visit adult volunteers with a confirmed or suspected diagnosis of chronic obstructive  pulmonary  disease  (COPD). The ground truth airflow limitation severity of each subject was determined by applying the Global Initiative  for  Chronic  Obstructive Lung Disease (GOLD) staging criteria to the subject's spirometry results. Spirometry was performed in a pulmonary function test laboratory under the supervision of trained clinical staff. Separately, the subjects' respiratory signal was measured during quiet breathing, and a classification model was built to infer the subjects' level of airflow limitation from the measured respiratory signal. The classification  model was evaluated against the ground truth using leave-one-out testing.

Results: Severity of airway obstruction was classified as either mild/moderate or severe/very severe with an accuracy of $96.4\%$. 

Conclusion: Tidal breathing parameters that are measured with a wearable device can be used to distinguish between different levels of airflow limitation in COPD patients. 



\end{abstract}


\section{Introduction}


Early detection and treatment of chronic obstructive pulmonary disease (COPD) exacerbations is critical to improving patients' quality of life. And frequent assessment of lung function could help with early detection: for instance, patients experience a significant drop in forced expiratory volume in 1 second (\fev) up to 2 weeks before an exacerbation \cite{watz2018spirometric}. Unfortunately, the state-of-the-art for measuring lung function at home is hand-held spirometry, a technique that is hindered by low patient adherence and poor patient technique in executing the spirometric breathing maneuvers \cite{cruz2014home}.

\begin{figure}[t]
\begin{center}
\includegraphics[scale=0.5]{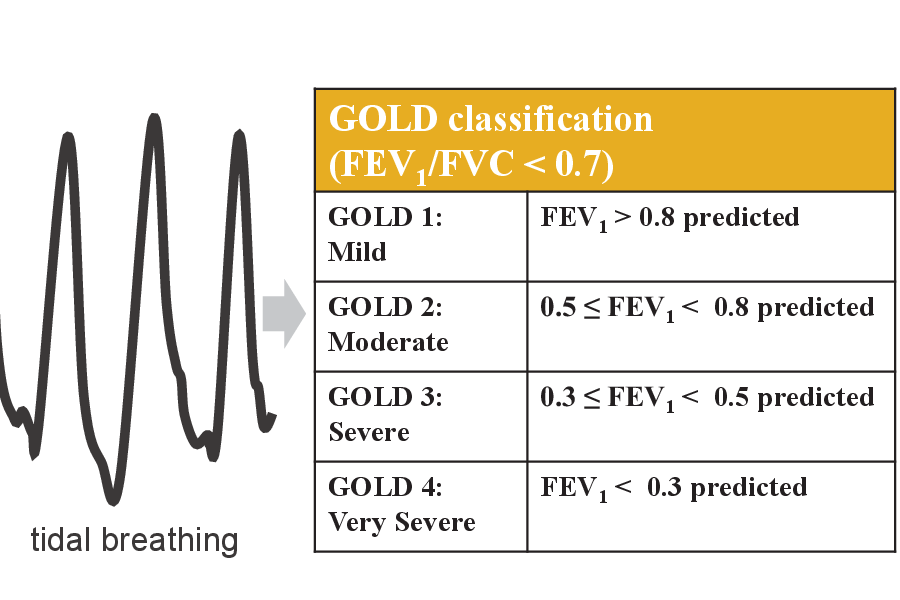}
\end{center}
\vspace{-5pt}
\caption{The objective of the proposed algorithm is to process the respiratory signal measured during tidal  breathing and produce a classification of airflow limitation severity that corresponds to one of the GOLD staging criteria \cite{singh2019global}.}
\label{figure:oneGwords}
\end{figure}

To address this problem, many studies over the past few decades have focused on monitoring COPD patients with mobile and wearable technology. Various approaches have been explored, from monitoring of physical activity \cite{sievi2017accelerometer, kawagoshi2013quantitative}, to accelerometer-based assessment of cardiorespiratory function \cite{naranjo2018smart, rahman2018instantrr, liaqat2019wearbreathing}. These previous studies have generally demonstrated that wearable sensor data can be used to distinguish between COPD patients and healthy controls. But they fall short in elucidating whether wearable technology can produce clinically useful information for COPD management. Specifically, none of the aforementioned studies has clarified whether a COPD patient's \underline{\textbf{severity level of lung airway obstruction}}---as classified by the Global Initiative for Chronic Obstructive Lung Disease (GOLD) \cite{singh2019global}---can be inferred from wearable sensors.




\begin{figure*}[t]
\begin{center}
\includegraphics[scale=0.52]{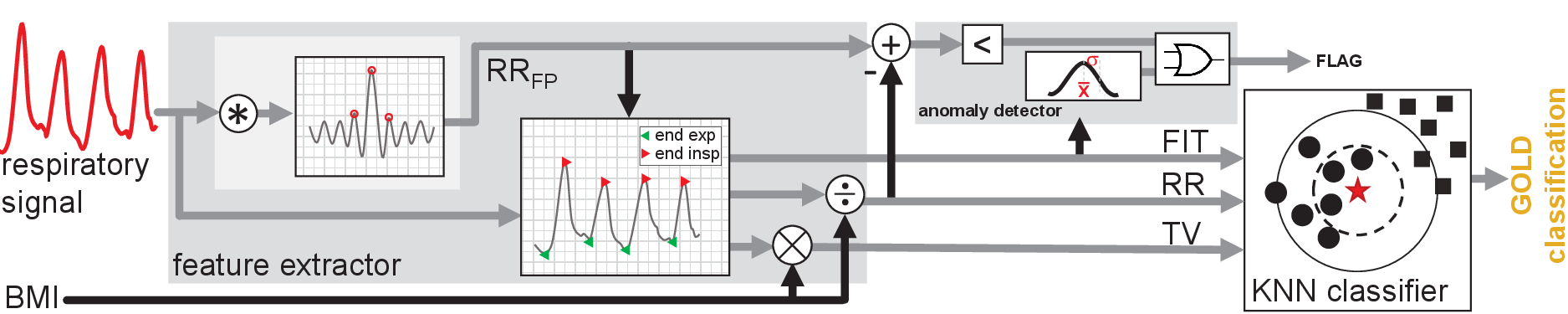}
\end{center}
\vspace{-10pt}
\caption{Block diagram of proposed algorithm. The  feature  extractor  processes  20  second  windows of  respiratory  data  at  a  time, to produce fractional inspiratory time (\fit),  respiratory  rate (\rr)  and  the  tidal volume (\tv). An anomaly detector throws a flag if the extracted features are outside the normal expected ranges (see text). Finally, a k-nearest neighbour classifier infers the GOLD classification airflow limitation level from the extracted features.}
\label{figure:blockDiagram}
\end{figure*}

A few researchers have gone beyond merely distinguishing between COPD patients and healthy controls, and have instead introduced new approaches to infer severity of airway obstruction from non-spirometric data \cite{bellos2013identification, ying2016gold, moghadas2017lung}. Unfortunately, these alternatives to spirometry depend on data that cannot easily be measured with a wearable device. Thus, these novel tools for estimating COPD severity are unsuitable for continuous, at-home patient monitoring. 

Unless wearable sensor data can be translated into a usable, clinical-standard rating of lung airway obstruction, these emerging technologies will have a limited impact on remote COPD management  \cite{piwek2016rise}. To tackle this problem, we have developed an algorithm (\figs{oneGwords}{blockDiagram}) that can infer a patient's GOLD classification level of lung airway obstruction from the respiratory signal measured during tidal (i.e. ``quiet'') breathing. The respiratory signal can easily be measured via accelerometry \cite{phan2008estimation} or cardiac-derived techniques \cite{7997854, 6497470, 4811954}, using unobtrusive wearable devices like tags, earphones, rings or smartwatches \cite{holt2018ambulatory, castaneda2018review, gaurav2016cuff, strik2020validating, kinnunen2020feasible}. Our proposed algorithm is a step towards bridging the gap between wearable sensor data and clinically relevant information for COPD management.

\section{Effect of Airway Obstruction on Tidal Breathing}
\label{section:efx}
The strict diagnosis of COPD requires compatible symptoms plus spirometry that demonstrates a ratio of  forced expiratory volume in 1 second to forced vital capacity (\fevf) that is less than  $70 \%$. Once a COPD diagnosis is made, effective management of the disease depends on monitoring the severity of airflow limitation, which is derived from the percentage predicted forced expiratory volume (\fev). Unfortunately, \fev~measurement requires active user engagement, is dependent on the patient's skill and motivation level, and is generally unreliable unless performed under skilled clinical supervision \cite{cruz2014home}.

As an alternative to spirometry, we propose to measure and analyze tidal, or ``quiet'' breathing, since the same pathology that produces abnormal \fev~values will also affect the patient's respiratory pattern when at rest. Following is a brief discussion of how airway obstruction affects various aspects of the normal respiratory pattern.

\subsection{Expiratory Phase Duration}
\label{section:exp}
COPD has two contributory pathologies, chronic bronchitis and emphysema, and most patients with COPD have both pathologies. Chronic bronchitis causes inflammation and excess mucus production in the lungs’ airways, resulting in impeded airflow. When breathing in (that is, during the ``inspiratory phase'' of respiration), COPD patients can overcome this airway obstruction by increasing the work of the diaphragm and intercostal muscles and breathing in more forcefully. Airway diameter also increases slightly in inspiration as pleural pressure becomes more negative than the atmosphere. Breathing out (during the ``expiratory phase'') is a more passive process than breathing in, relying on elastic recoil rather than muscular action during normal respiration. So, with no active compensatory mechanism to counter it, airway obstruction will slow down the rate at which patients can breathe out and thus prolong the expiratory phase \cite{malhotra2018research}. 

Small airways in the emphysematous lung can also collapse in exhalation, a major driver of airflow obstruction in patients with COPD, leading to slow expiratory phase. Some patients with emphysema will utilize pursed lip breathing to facilitate maximal exhalation, which also contributes to a long expiratory phase \cite{de2015pursed}.



\subsection{Tidal Volume}
\label{section:tidal}
At the end of the expiratory phase of tidal breathing, there is an equilibrium between the elastic recoil pressure of the chest wall and that of the lungs. This pressure equilibrium will always leave the lungs partially inflated with some residual amount of air, or functional residual capacity (FRC). But COPD patients experience a loss of lung elastic recoil, and this causes an increase in FRC \cite{dube2015clinical, ferguson2006does}. As such, the amount of additional air that COPD patients can inhale during  tidal breathing (known as ``tidal volume'') is decreased  \cite{larsson2007aspects}.


\subsection{Respiratory Rate}
In addition to diminished elasticity, the air sacs of COPD patients suffer wall damage, which reduces their surface area and decreases diffusion of oxygen into, and carbon dioxide out of, the bloodstream. COPD patients typically overcome this gas exchange inefficiency by breathing faster to increase their lung ventilation \cite{aliverti2014mechanics, loring2009pulmonary}. So, worsening COPD is often marked by an increased resting respiratory rate.


\section{Proposed Algorithm}

As the block diagram of \fig{blockDiagram} shows, our algorithm for inferring COPD severity receives two inputs: raw respiratory data and body-mass index (BMI). The output is a COPD severity classification that corresponds to GOLD staging criteria \cite{singh2019global}. In this section, we discuss the main components of the algorithm, which comprises a feature extractor, an anomaly detector, and a k-nearest neighbours (KNN) classifier. 

\subsection{Feature Extractor}
\begin{figure}
\begin{center}
\includegraphics[scale=0.54]{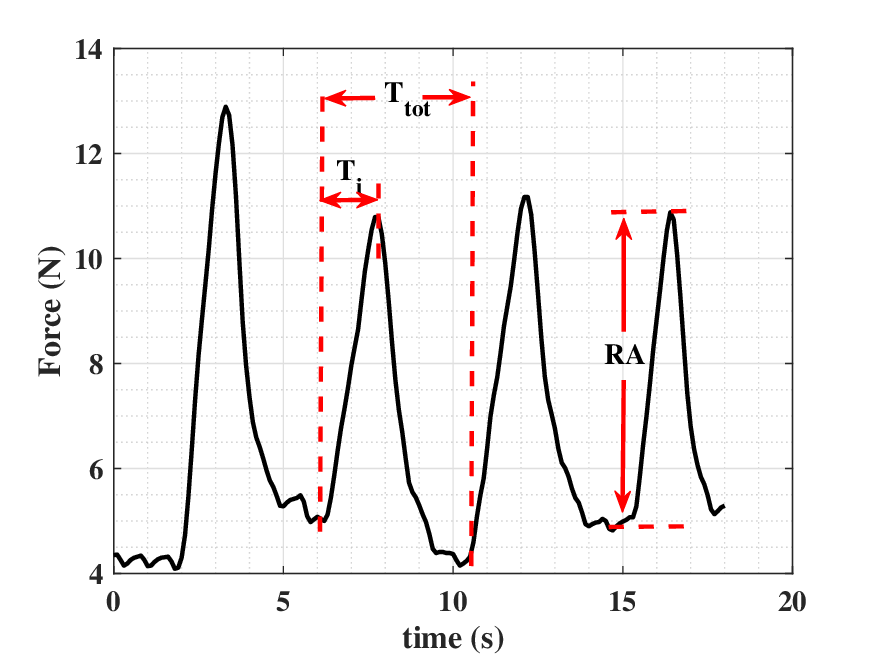}
\end{center}
\vspace{-10pt}
\caption{A typical respiratory signal. The inspiratory phase lasts $T_{\rm i}$ seconds, and the full respiratory cycle (inspiratory and expiratory phases) lasts $T_{\rm tot}$ seconds. The amplitude of the signal is measured from trough to peak as $RA$ Newtons.}
\label{figure:respsignal}
\end{figure}

The feature extractor processes 20 second windows of respiratory data at a time.  For each window, the feature extractor first calculates a first-pass estimate of respiratory period (\rr$_{\rm FP}$ in \fig{blockDiagram}) as the average distance between peaks of the autocorrelated signal. Next, the algorithm identifies the local maxima and local minima, which correspond to the end inspiratory and expiratory points, respectively, of each breathing cycle. The algorithm avoids spurious end points by ignoring any maxima (or minima) that are separated by less than 0.6 of the first-pass respiratory period. The algorithm also ignores any maxima (minima) with a height (depth) that is less than 30 \% of the highest (lowest) point in the 20 s window. From the inspiratory and expiratory end points, the feature extractor calculates the fractional inspiratory time, respiratory period and the respiratory amplitude for each breathing cycle (see \fig{respsignal}). 


Fractional inspiratory time (\fit) is the ratio of the inspiratory time to the total respiratory period. Referring to \fig{respsignal}, we calculate this as \fit~$= T_{\rm i}/T_{\rm tot}$. Note that the quantity $1-$\fit~is a measure of the expiratory phase duration, normalized to the respiratory period.

The respiratory rate is the average number of breaths taken per minute, calculated as $\rr=60/({\rm BMI}\cdot T_{\rm tot})$. We normalize the respiratory rate with body-mass index (BMI), because they are significantly correlated \cite{chlif2009effects, littleton2012impact}.

The respiratory amplitude, $RA$, is the difference in force between an adjacent trough-peak pair (see \fig{respsignal}). This quantity provides an approximation of the tidal volume as follows. The respiratory  amplitude is proportional to the change in radius of the chest as it expands during the inspiratory phase. Since BMI is proportional to chest area \cite{ghoshhajra2011direct}, we can estimate the tidal volume (modulo constants of proportionality) as \tv$=RA\cdot{\rm BMI}$.

\subsection{Anomaly Detector} 
Wearable device data are prone to being corrupted by the everyday disturbances encountered in a home environment \cite{kang2018recent}. So, our proposed algorithm includes an anomaly detector to flag and discard potentially noisy data. The anomaly detector throws a flag if either of the following conditions is met for a given 20 second window: (1) the ratio of the \fit~mean to standard deviation is less than 3.33; or (2) there is more than a 10 \% difference in the respiratory rate estimated from autocorrelation versus that estimated directly from the raw respiratory signal. 


\subsection{KNN Classifier} 
The k-nearest neighbor (KNN) classifier infers the patient's COPD severity level from the \fit, \rr~and \tv~features. When presented with the feature vector of a previously-unseen patient, the classifier will assign the severity level label (GOLD stage 1, 2, 3 or 4, corresponding to `mild', `moderate', `severe' or `very severe' airflow limitation) that is most similar to the vector's closest 7 neighbors. The metric used to identify the closest neighbors is the cosine distance measure, which is defined for two vectors $x_1$ and $x_2$ as
\begin{equation}
d(x_1, x_2) = \frac{x_1\cdot x_2}{\lVert x_1 \rVert \cdot \lVert x_2 \rVert}.
\end{equation}


\section{Experimental Method}
To evaluate the performance of our algorithm, we collected and analyzed data from patients at the pulmonary clinic of the Dartmouth Hitchcock Medical Center (Lebanon, NH). We obtained informed consent from each subject before their inclusion in the study. All aspects of the study protocol were reviewed and approved by the Dartmouth College Institutional Review Board (Committee for the Protection of Human Subjects-Dartmouth; Protocol Number: 00028641).
\subsection{Subjects}
Twenty-five adults, 10 men and 15 women, with either suspected or previously-diagnosed COPD, and scheduled to receive a physician-ordered spirometry test, were recruited for this study. We were unable to collect more data before the onset of the Covid-19 pandemic in March 2020 halted all routine spirometry testing, due to the risk of aerosolized droplets \cite{helgeson2020aerosol}. A summary of the subjects' spirometry test results and their anthropomorphic data is listed in Table \ref{tab:subjects}.

\begin{table}
    \centering
    \begin{tabular}{|l|r|} \hline 
    & mean $\pm$ std \\ \hline 
    Age (y) & $67.6 \pm 11.6$ \\ \hline
    Height (cm) & $165.3 \pm 10.7$ \\ \hline
    Weight (kg) & $73.6 \pm 17.1$ \\ \hline
    BMI (kg/m$^2$) & $27.0 \pm 6.08$ \\ \hline
    \fevf~& $0.52 \pm 0.16$ \\ \hline
    \fev~(L) & $1.50 \pm 0.81$ \\ \hline
    \fvc~(L) & $2.68 \pm 1.09$ \\ \hline
    \end{tabular}
    \caption{Summary of anthropomorphic data and spirometry test results of the subjects recruited for this study (n=$25$, 10 men).}
    \label{tab:subjects}
\end{table}

\subsection{Data Collection}
First, each subject donned a Go Direct Respiration Belt (Vernier Software \& Technology, Beaverton, OR) and breathed quietly while seated. The Respiration Belt is an instrumented strap of fabric that is worn just below the breast bone. It has an embedded accelerometer that measures the force produced by the chest as the subject breathes in and out. \figur{respsignal} is a typical example of the resulting tidal breathing signal waveform that we collected from each subject. Each subject produced 15-20 breaths (approximately 1 minute of data collection).

Next, we collected spirometry data to establish the ground truth of the patient's level of COPD severity. Each subject performed standard spirometry with a PC-based spirometer under the supervision of trained clinical staff. The subject performed three spirometery maneuvers that were acceptable and reproducible according to the American Thoracic Society criteria. We collected the subject's best values of \fevf, \fev~and \fvc~for subsequent data analysis. 


\subsection{Data Analysis}
We generated a correlation matrix to study the relationship between the tidal breathing parameters (fractional inspiratory time, respiratory rate, and estimated tidal volume) and the spirometric variables (\fevf, \fev, \fvc). We further explored the relationships between these two sets of parameters via linear regression analysis.

To accommodate the limits on data collection caused by the Covid-19 pandemic, we employed a number of techniques to rigorously evaluate the performance of our algorithm. The low total number of subjects enrolled in the study meant that there was an insufficient number of subjects to measure the classification accuracy for every COPD severity level with adequate power \cite{sim2005kappa}. We addressed this problem by grouping the patients into two classes: mild/moderate and severe/very severe. Further, we calculated the confidence intervals of our measured sensitivity and specificity. Also, we tested our algorithm using leave-one-out cross validation as follows. We held out data from each subject in turn, and built a KNN model on the remaining data, which was augmented via synthetic minority oversampling \cite{chawla2002smote}. The resulting model was evaluated on the left-out data sample. Results were aggregated over all 25 folds in order to calculate sensitivity and specificity values.

\section{Results and Discussion}



\begin{figure}
\begin{center}
\includegraphics[scale=0.54]{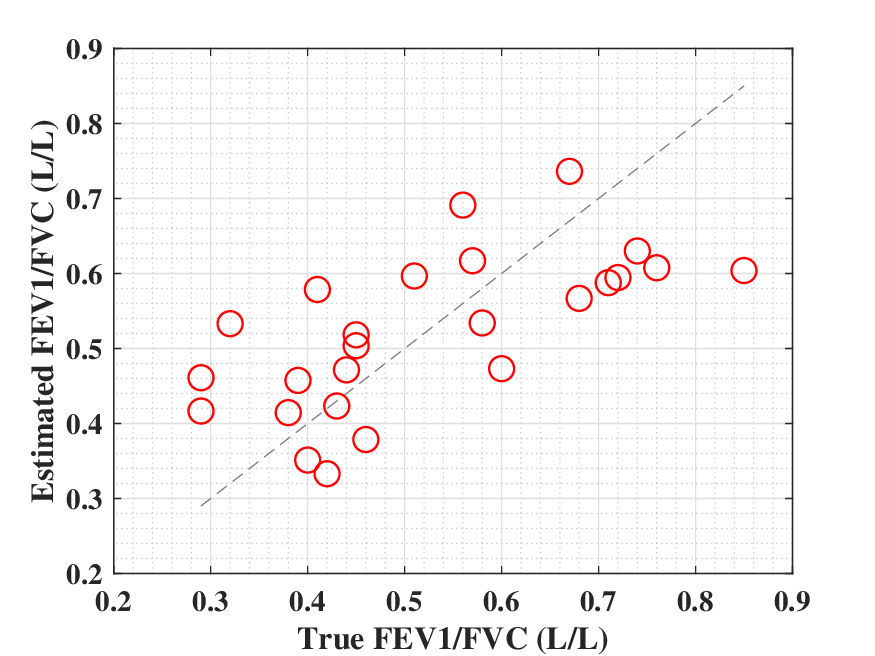}
\end{center}
\vspace{-10pt}
\caption{\fevf~regression model predictions versus true values. The dashed gray curve is the line of equality. Coefficient of determination $R^2=0.435$, with $p=0.002$.}
\label{figure:fevfvcregress}
\end{figure}

Table \ref{tab:crossco}, \figs{fevfvcregress}{fevregress} show the correlation between tidal breathing parameters and spirometric variables. From Table \ref{tab:crossco}, \fev~and \fevf~are most strongly correlated with \fit~($R^2=0.295$, $p=0.005$ and $R^2=0.274$, $p=0.007$ respectively), while \fvc~is most strongly correlated with tidal volume ($R^2=0.329$, $p=0.003$). These results are in line with the observation (see \sect{efx}) that airway obstruction is a factor that determines both \fit~and the spirometer parameters, \fevf~and \fev. Similarly, \fvc~showed the strongest correlation with tidal volume, due to the effect of lung elastic recoil on both these measures.

\begin{table}
    \centering
    \begin{tabular}{|c|c|c|c|} \hline 
    & \textbf{\fevf}   & \textbf{\fev} & \textbf{\fvc}\\ \hline \hline
    \textbf{\fit}  & 0.274 & 0.295 & 0.135 \\
     &  \footnotesize{$p=0.007$} & \footnotesize{$p=0.005$} &  \footnotesize{$p=0.070$} \\ \hline
    \textbf{\rr} & 0.129 & 0.08 & 0.0125 \\
     &  \footnotesize{$p=0.078$} &  \footnotesize{$p=0.171$} &  \footnotesize{$p=0.595$} \\ \hline
    \textbf{\tv} & 0.060 & 0.317 & 0.329 \\
     &  \footnotesize{$p=0.238$} &  \footnotesize{$p=0.003$} &  \footnotesize{$p=0.003$} \\ \hline
    \end{tabular}
    \caption{Matrix of correlations between tidal breathing parameters and spirometric variables.}
    \label{tab:crossco}
\end{table}


\begin{figure}
\begin{center}
\includegraphics[scale=0.54]{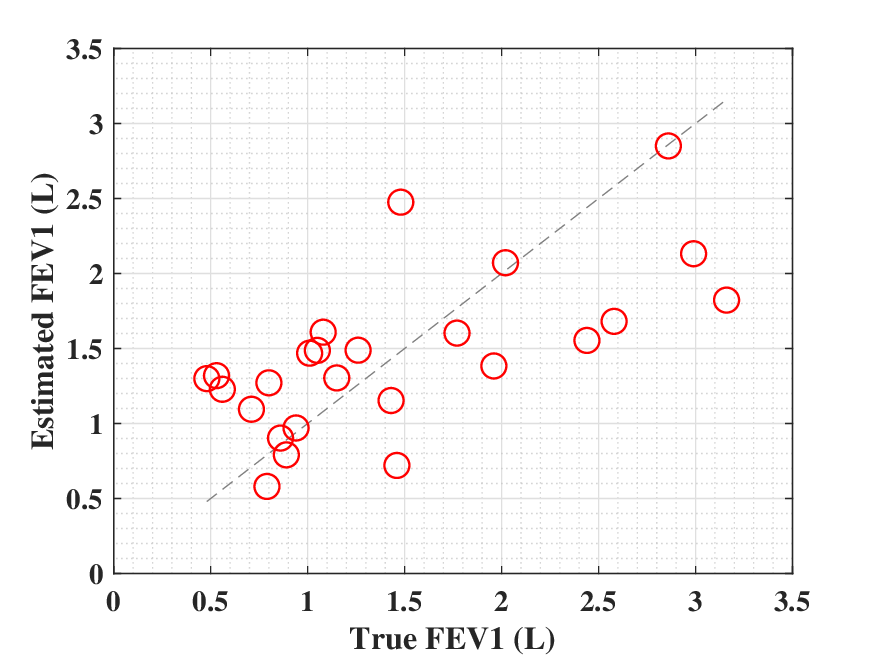}
\end{center}
\vspace{-10pt}
\caption{\fev~regression model predictions versus true values. The dashed gray curve is the line of equality. Coefficient of determination $R^2=0.427$, with $p=0.002$.}
\label{figure:fevregress}
\end{figure}

\begin{figure}
\begin{center}
\includegraphics[scale=0.54]{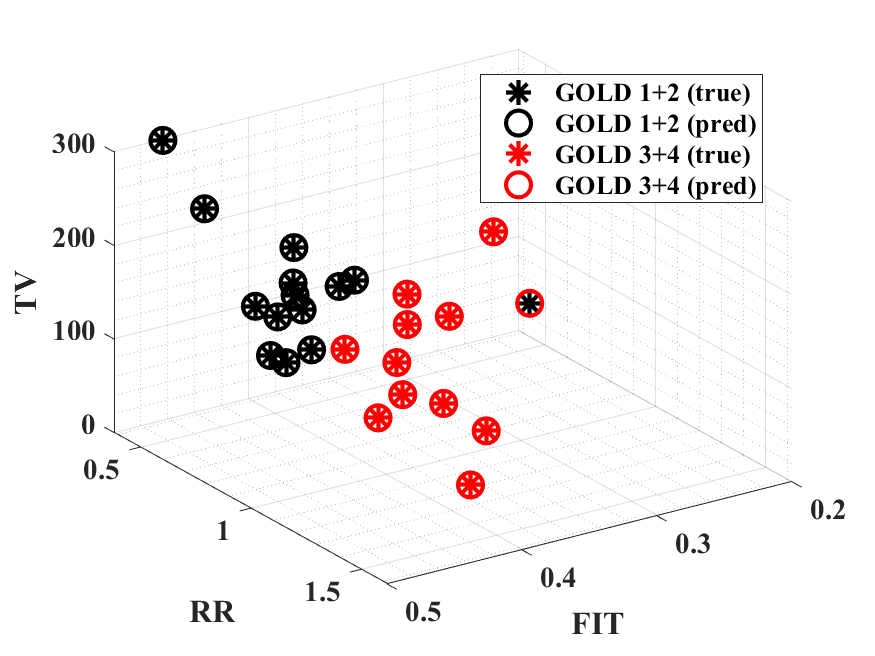}
\end{center}
\vspace{-20pt}
\caption{Scatter plot of airflow limitation severity class labels versus tidal breathing parameter features. Black markers indicate mild or moderate airway obstruction, while red markers indicate severe or very severe obstruction. The asterisks are the true class labels, and the circles are the classifier's inferred output.}
\label{figure:fevclass}
\end{figure}

The performance of the severity classification model is summarized in Table \ref{tab:severityCoarse}, \figst{fevclass}{fevconfusion}{roc}. The point estimate values for sensitivity and specificity are $100\%$
and $92.8\%$, respectively. Also, the area under the receiver operating characteristic curve (AUC) is $0.93$. Our algorithm can reliably distinguish between COPD GOLD stages 1+2 and GOLD stages 3+4. 

Table \ref{tab:comparison} compares our work to other studies in the literature. The different approaches all yield generally the same level of classification accuracy, but our work differs in one critical aspect: while others depend on extensive medical records \cite{moghadas2017lung, ying2016gold}, questionnaires \cite{bellos2013identification} or complex protocols and equipment \cite{altan2020chronic} to make inferences, our approach only requires the user's BMI and respiratory signal. For most patients, BMI need only be measured periodically, during the annual check-up. And respiratory signal can be continuously measured with any number of wearable devices \cite{holt2018ambulatory, castaneda2018review, gaurav2016cuff, strik2020validating, kinnunen2020feasible}.

\begin{figure}
\begin{center}
\includegraphics[scale=0.55]{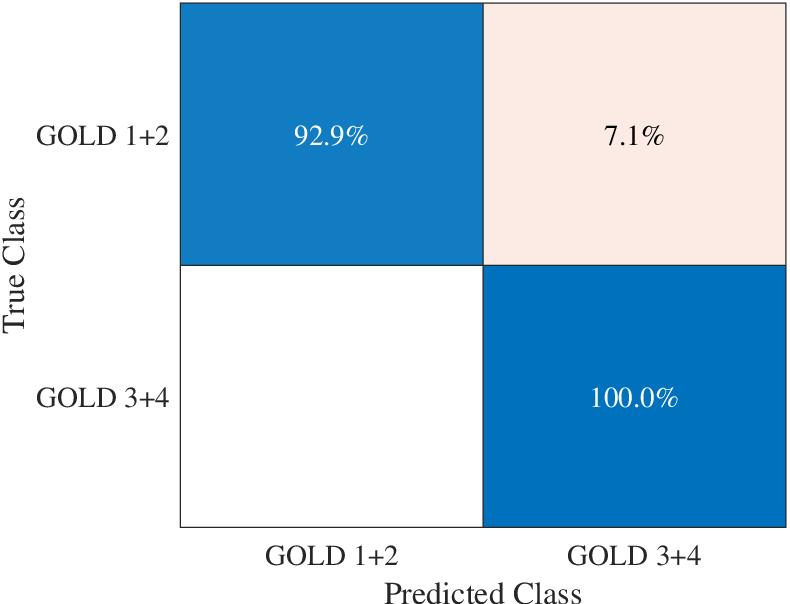}
\end{center}
\vspace{-10pt}
\caption{Confusion matrix for airflow limitation severity inference.}
\label{figure:fevconfusion}
\end{figure}

\begin{figure}
\begin{center}
\includegraphics[scale=0.55]{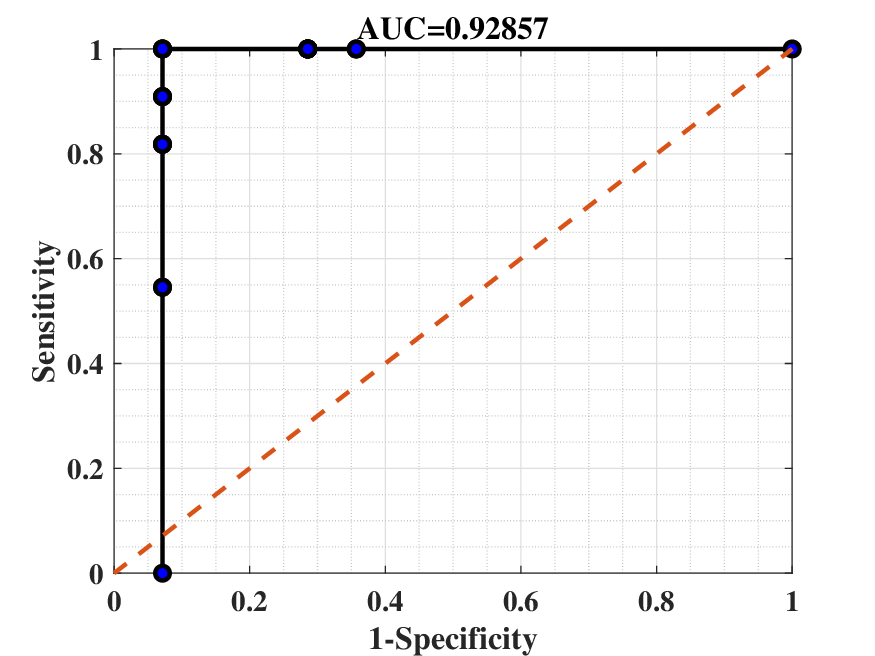}
\end{center}
\caption{Receiver operating characteristic curve for airflow limitation severity inference, where `severe/very severe' is the positive class.}
\label{figure:roc}
\end{figure}



\begin{table}
    \centering
    \begin{tabular}{|c|c|c|c|} \hline 
       & \textbf {Estimate} & \textbf{80\% CL} & \textbf{95\% CL} \\ \hline \hline
    \textbf{Sens (\%)} & 100 & [81.1, 100] & [71.5, 100] \\ \hline
    \textbf{Spec (\%)} & 92.8 & [74.9, 99.2] & [66.1, 99.8] \\ \hline
    \end{tabular}
    \caption{Performance of classifier for stratification of airway obstruction severity, where GOLD stage 3 and 4 (severe/very severe) is the positive class, and GOLD stage 1 and 2 (mild/moderate) is  the negative class. The point estimate is reported, along with the 80\% and 95\% confidence limits.}
    \label{tab:severityCoarse}
\end{table}


\begin{table*}[th]
    \centering
    \begin{tabular}{|c|c|l|c|} \hline 
       & \textbf {Accuracy (\%)} & \textbf{Source Data}  & \textbf{Wearable?} \\ \hline \hline
    \textbf{\cite{moghadas2017lung}} & 84 & Computed tomography (CT) scans & N \\ \hline
    \textbf{\cite{bellos2013identification}} & 80 & Blood glucose, environmental humidity, questionnaires, etc.  & N \\ \hline
    \textbf{\cite{ying2016gold}} & 97 & Patient medical records  & N \\ \hline
    \textbf{\cite{altan2020chronic}} & 94 & 12-point lung auscultation  & N \\ \hline
    \textbf{This work} & 96 & Respiratory signal  & Y \\ \hline
    \end{tabular}
    \caption{Comparison of different approaches for inferring severity of airflow limitation in  COPD patients.}
    \label{tab:comparison}
\end{table*}



\section{Conclusion}
In this paper, we presented an algorithm to infer the severity of a COPD patient's airflow limitation from tidal breathing data that was collected by a wearable device. We evaluated the algorithm on adult subjects  with a confirmed or suspected diagnosis of chronic obstructive  pulmonary  disease (COPD), and we confirmed that the algorithm is able to stratify the severity of patients' airway obstruction. The results of this study provide a strong premise for further data collection and exploring the concept of wearable devices for lung health monitoring.

\section{Acknowledgements}
This work was supported in part by the Munck-Pfefferkorn Education and Research Fund.
We would like to thank Jonathan Toledo and Orkan Sezer of Dartmouth College and Henry Berube, Jeffrey Vonada and Jennifer Hilton-Hancock of Dartmouth Hitchcock Medical Center for their invaluable support with data collection. The views and conclusions contained in this document are those of the authors and should not be interpreted as necessarily representing the official policies, either expressed or implied, of the sponsors.
\printbibliography
\end{document}